\documentclass[aps,preprint,unsortedaddress,floats]{revtex4}
\usepackage{amsmath,amsfonts,graphicx}

\begin{document}
\bibliographystyle{revtex}
\draft

\textwidth 16cm \textheight 23cm \topmargin -1cm \oddsidemargin
0cm \evensidemargin 0cm

\title{Electron acceleration and heating in collisionless magnetic reconnection}

\author{Paolo Ricci$^{a,b}$}
\author{Giovanni Lapenta$^{a,c}$}\email{lapenta@lanl.gov}
\author{J.U. Brackbill$^{c}$} \email{jub@lanl.gov}

\affiliation{a) Istituto Nazionale per la Fisica della Materia
(INFM), Unit\`a del Politecnico di Torino, Corso Duca degli
Abruzzi 24 - 10129 Torino, Italy}

\affiliation{b) Dipartimento di Energetica, Politecnico di Torino,
Italy}

\affiliation{c) Theoretical Division, Los Alamos National
Laboratory, Los Alamos NM 87545}

\date{\today}

\begin{abstract}

We discuss electron acceleration and heating during collisionless
magnetic reconnection by using the results of implicit kinetic
simulations of Harris current sheets. We consider and compare
electron dynamics in plasmas with different $\beta$ values and
perform simulations up to the physical mass ratio. We analyze the
typical trajectory of electrons passing through the reconnection
region, we study the electron velocity, focusing on the
out-of-plane velocity, and we discuss the electron heating along
the in-plane and out-of-plane directions.

\end{abstract}

\maketitle

\section{INTRODUCTION}

Collisionless magnetic reconnection plays an important role in
energetically active processes in plasmas \cite{Biskamp2000,
Priest2000}. Magnetic reconnection takes place in plasmas
characterized by different values of $\beta$. Theoretical,
observational, and experimental results show that reconnection is
present in the geomagnetic tail \cite{Oieoroset2001}, where local
$\beta >> 1$; in the Earth's magnetopause \cite{Nishida1978},
where $\beta \approx 1$; in laboratory \cite{Syrovatsky1973,
Gekelman1991, Yamada1999, Egedal2001, Furno2003, Taylor1996}, in
the solar corona plasma \cite{Priest1982}, and in astrophysical
plasmas, such as extragalactic jets \cite{Romanova1992,
Blackman1996, Lesh1998} and flares in Active Galactic Nuclei (AGN)
\cite{Lesch1997}, where $\beta \leq 1$.

During magnetic reconnection, magnetic energy is converted into
kinetic and thermal energy of electrons and ions. In fact,
electron heating and acceleration are signatures of magnetic
reconnection. In the magnetotail, bursts of energetic electrons
have been attributed to reconnection \cite{Terasawa1976,
Baker1976, Baker1977} and there have been recent direct
measurements of electron acceleration during magnetic reconnection
\cite{Oieoroset2002}. Production of runaway electrons during
sawtooth instabilities and disruptions is associated with magnetic
reconnection in tokamaks \cite{Helander2002}. In solar flares,
x-ray observations indicate that a large fraction of the total
energy is released in accelerated electrons \cite{Lin1971,
Lin1976, Gibson2002}. The observed synchroton radiation in
extragalatic jets is thought to be generated by reacceleration or
in-situ acceleration of electrons due to magnetic reconnection
\cite{Romanova1992, Lesh1998}. It has been proposed that the
detection of hard x-ray and $\gamma$-ray from AGN is due to the
presence of electrons accelerated by magnetic reconnection
\cite{Lesch1997}.

Electron dynamics in the reconnection region have been studied
using analytical arguments \cite{Speiser1965, Zelenyi1990,
Litvinenko1996, Litvinenko1997}, test particle theory
\cite{Wagner1980, Sato1982, Scholer1987, Deeg1991, Vekstein1997,
Birn2000}, self-consistent fluid simulations \cite{Nodes2003}, and
kinetic simulations \cite{Loboeuf1982, Horiuchi1997, Hoshino2001}.

The aim of the present paper is to study the electron dynamics
near the reconnection region with self-consistent kinetic
simulations of high and low $\beta$ plasmas. The plasma $\beta$ is
varied from very large ($\beta >> 1$) to small ($\beta<1$) values
and systems are simulated with an ion/electron mass ratio up to
the physical value ($m_i/m_e=1836$). We consider two-dimensional
reconnection in Harris current sheet configurations
\cite{Harris1962}, triggered by an initial perturbation
\cite{Birn2001}. We introduce a guide field to reduce the plasma
$\beta$ and eliminate the null field region at the current sheet.
To perform kinetic simulations, we use CELESTE3D
\cite{Brackbill1985, Vu1992, Ricci2002a}, an implicit
Particle-in-Cell (PIC) code, which models both kinetic ions and
electrons while allowing simulations with higher mass ratios.

We show that both the plasma $\beta$ and the mass ratio strongly
affect the electron dynamics. In low-$\beta$ plasmas, the electron
meandering orbits present in high $\beta$ plasmas disappear. We
focus on the out-of-plane electron velocity, which remains
localized in the high-$\beta$ case, and which is sizable also far
from the reconnection region in low $\beta$ plasmas
\cite{Horiuchi1997}. A strong influence of the mass ratio on the
out-of-plane velocity is shown by the simulations and the relevant
scaling law is deduced. We show that the heating process is
non-isotropic in presence of a guide field; in particular, the
particles are preferably heated in the out-of-plane direction.
This anisotropy contributes to the break-up of the frozen-in
condition for electrons and allows reconnection to happen.

The paper is organized as follows. Section II describes the
physical system and the simulations. Section III presents the
results of the simulations, focusing on the typical trajectory of
the electrons that pass through the reconnection region and the
evolution of the electron fluid velocity and temperature during
reconnection.

\section{PHYSICAL SYSTEM}

We consider a two-dimensional Harris current sheet in the $(x,z)$
plane \cite{Harris1962}, with an initial magnetic field given by
\begin{equation}
\mathbf{B}_{0}(z)=B_0 \tanh(z/\lambda) \mathbf{e}_x+ B_{y0}
\mathbf{e}_y
\end{equation}
and plasma particle distribution functions for the species $s$
($s=e,i$) by
\begin{eqnarray}
f_{0,s}(z, \mathbf{v})& = & n_0 \, \hbox{sech} ^2(z/ \lambda)
\left(\frac{m_s}{2 \pi k_B T_s} \right)^{3/2} \exp \left\{ -
\frac{m_s}{2 k_B T_s} \left[ v_x^2+(v_y-V_s)^2+v_z^2 \right]
\right\} \nonumber \\
& + & n_b \left(\frac{m_s}{2 \pi k_B T_s} \right) ^{3/2} \exp
\left( - \frac{m_s v^2}{2 k_B T_s} \right)
\end{eqnarray}

We use the same physical parameters as the GEM challenge
\cite{Birn2001}. The temperature ratio is $T_i/T_e=5$, the current
sheet thickness is $\lambda=0.5 d_{i}$, the background density is
$n_b=0.2 n_0$, and the ion drift velocity in the $y$ direction is
$V_{i0}=1.67V_A$, where $V_A$ is the Alfv\'en velocity, and
$V_{e0}/V_{i0}=-T_{e0}/T_{i0}$. The ion inertial length,
$d_i=c/\omega_{pi}$, is defined using the density $n_0$. We apply
periodic boundary conditions in the $x$ direction and perfect
conductors in the $z$ direction.

The standard GEM challenge parameters model reconnection in high
$\beta$ plasmas. To model low $\beta$ plasmas, it is possible to
consider either a entirely different equilibrium
\cite{Nishimura2003}, or one may introduce a guide field in the
standard Harris sheet equilibrium. Herein, we follow the second
approach and we introduce a guide field $B_y=B_{y0}$, with a
spatially constant value at $t=0$. The simulations are performed
with different mass ratios, ranging from $m_i/m_e=25$ (standard
GEM mass ratio) to the physical mass ratio for hydrogen,
$m_i/m_e=1836$. Following {\it Birn et al.} \cite{Birn2001}, the
Harris equilibrium is modified by introducing an initial flux
perturbation in the form
\begin{equation}
A_y=-A_{y0} \cos (2 \pi x / L_x) \cos (\pi z /L_z)
\end{equation}
with $A_{y0}=0.1 B_0 c / \omega_{pi}$, which puts the system in
the non-linear regime from the beginning of the simulation.

The simulations shown in the present paper are performed using the
implicit PIC code CELESTE3D, which solves the full set of
Maxwell-Vlasov equations using the implicit moment method
\cite{Brackbill1985, Vu1992, Ricci2002a}. The implicit method
allows more rapid simulations on ion length and time scales than
are allowed with explicit methods, yet retains the kinetic effects
of both electrons and ions. In particular, the explicit time step
and grid spacing limits are replaced in implicit simulations by an
accuracy condition, $v_{th,e} \Delta t < \Delta x$, whose
principal effect is to determine how well energy is conserved. In
the simulations shown below, we typically choose $\omega_{ce}
\Delta t \approx 0.5$, $\Delta x /d_i=0.4$, and $\Delta
z/d_i=0.2$.

Previous work on magnetic reconnection performed by CELESTE3D have
proved that results from our implicit code match well the results
of explicit codes \cite{Ricci2002a}. Implicit simulations allow
one to model physical mass ratios, with which it is possible to
distinguish scaling laws associated with different break-up
mechanisms \cite{Ricci2002b}. CELESTE3D has also been employed in
a comprehensive study of the physics of fast magnetic
reconnection, in plasmas characterized by different $\beta$ values
\cite{Ricci2003}.

\section{RESULTS}

We have performed a set of simulations, using different mass
ratios ($m_i/m_e=25, 180, 1836$) and introducing different guide
fields: $B_{y0}=0$, with $\beta= \infty$ at the center of the
current sheet; $B_{y0}= B_{0}$, with $\beta=1.2$; and $B_{y0}=5
B_{0}$, with $\beta=0.048$. We note that a guide field changes
drastically the magnetic configuration of the system as the X
point is no longer a null-point, as it is in the $B_{y0}=0$ case.

The dynamics of magnetic reconnection in plasmas with different
values of $\beta$ have been pointed out and summarized in a
previous paper \cite{Ricci2003}. Figure 1 shows the reconnected
flux, $\Delta \Psi$, defined as the flux difference between the X
and the O points for all the simulations considered
\cite{Ricci2003}. Even though an initial perturbation is applied,
reconnection proceeds slowly during an initial transient phase
(which lasts approximatively until $t\omega_{ci} \approx 10$),
when the system adjust to the initial perturbation. Subsequently,
reconnection develops rapidly until the saturation level is
reached. Both the reconnection rate and the saturation level
decrease when the guide field is increased. All the simulations
show a similar evolution. The mechanism which breaks the electron
frozen-in condition is provided by the off-diagonal terms of the
electron pressure tensor for all the guide fields considered
\cite{Ricci2003, Hesse2002}. The reconnection rate is enhanced by
the whistler dynamics in high $\beta$ plasmas, and by the Kinetic
Alfv\'en Waves dynamics in low $\beta$ plasmas \cite{Ricci2003,
Biskamp1997a, Sonnerup1979, Terasawa1983, Kleva1995, Rogers2001},
provided that $\beta > m_e/m_i$. When $\beta < m_e/m_i$, fast
reconnection is not possible \cite{Biskamp1997b, Ottaviani1993}.

Theoretical results and kinetic simulations \cite{Ricci2002b,
Kuznetsova2000} show that with $B_{y0}=0$, electrons flow towards
the X point along the $z$ direction, where they are demagnetized
in a region corresponding to the meandering length, $d_{ze}$.
There they are accelerated by the reconnection electric field,
$E_y$, in the $y$ direction. The electrons are then diverted by
the $B_z$ field, gaining an outflow velocity in the $x$ direction,
and becoming remagnetized at the meandering length, $d_{xe}$. The
meandering lengths are defined as \cite{Ricci2002b,
Kuznetsova2000}

\begin{equation}
d_{xe}=\left[ \frac{m_e T_e}{e^2 (\partial B_z/ \partial x) ^2}
\right] ^{1/4},d_{ze}=\left[ \frac{m_e T_e}{e^2 ( \partial B_x/
\partial z) ^2} \right] ^{1/4}
\label{medis}
\end{equation}
while the maximum inflow and outflow velocities scale as
\cite{Ricci2002b, Kuznetsova2000}

\begin{equation}
v_{xe}=\left[ \frac{e^2 E^4_y}{4 m_e T_e (\partial B_z/ \partial
x) ^2} \right] ^{1/4},v_{ze}=\left[ \frac{e^2 E^4_y}{4 m_e T_e
(\partial B_x/ \partial z) ^2} \right] ^{1/4}
\label{vel}
\end{equation}

In the reconnection region, the electrons are unmagnetized and
follow complex meandering orbits, which result in a non-gyrotropic
electron distribution function and in off-diagonal terms of the
electron pressure \cite{Hoshino1998}.

When a guide field is introduced, the meandering orbits disappear.
Analytical estimates of the guide field at which this happens are
given by \cite{Litvinenko1996, Litvinenko1997}. Nevertheless, the
diagonal components of the electron pressure tensor are unequal,
which contributes to the presence of off-diagonal pressure terms
\cite{Ricci2003, Hesse2002}. These terms still constitute the
break-up mechanism of the frozen-in condition \cite{Ricci2003,
Hesse2002}. In the reconnection region, the electrons flow across
the field line in the $(x,z)$ plane, while performing a Larmor
motion around the out-of-plane magnetic field. Far from the
reconnection region, the guide field causes additional components
of the $\mathbf{E} \times \mathbf{B}$ force, which modify the ion
and electron motion and cause asymmetric plasma flow.

Below, we describe in detail the typical trajectory of an electron
passing through the reconnection region in high and low $\beta$
plasmas. Then, we focus on the electron fluid velocity, in
particular on the out-of-plane velocity. Finally, we consider the
electron distribution functions to evaluate the electron
temperature.

\subsection{Electron trajectory}

Figure 2 shows a typical trajectory of an electron passing through
the reconnection region in the case $B_{y0}=0$, and Fig. 3 shows
the history of its velocity and kinetic energy.

In Fig. 2, the initial position of the particle is denoted by a
plus sign, and its position at selected time steps by circles. 'X'
marks the position of the X point. Note that periodic boundary
conditions are applied in the $x$ direction, as the behavior of
the trajectory shows (the particle exits from the left and
re-enters from the right). At the beginning, the electron is tied
to a magnetic field line (magnetic field lines run mainly along
the $x$ direction). Near the reconnection point, the electron
decouples from the magnetic field and moves along the $z$
direction, reaching the X point at $t\omega_{ci} \approx 15$. The
particle trajectory is meandering in the unmagnetized region. The
outflow from the reconnection region takes place as soon as the
electron reaches a region with stronger magnetic field. Then, the
electron couples again to a magnetic line surrounding the O point,
and starts again its gyration orbit around it.

In Fig. 3, all components of the particle velocity and the kinetic
energy are plotted as a function of time. Initially, the electron
is flowing along the $x$ direction, with a Larmor motion mostly in
the $(y,z)$ plane, which is responsible for the high frequency
oscillations of the velocity (the magnetic field line is mostly
directed along $x$). During this phase, the electron kinetic
energy is almost conserved. When the electron decouples from the
magnetic field line as it crosses the reconnection region, it is
accelerated by the reconnection electric field in the $y$
direction. This acceleration transfers magnetic field energy to
electrons during magnetic reconnection. As Fig. 3 shows, while the
electron is unmagnetized, the kinetic energy of the electron
increases remarkably, showing high frequency oscillations due to
the acceleration and de-acceleration by the electric field. When
the electron leaves the reconnection region and again couples to
the magnetic field, motion in $y$ becomes Larmor motion with a
bigger radius, and the velocity directed along $y$ gained in the
reconnection region is lost. The particle couples to a magnetic
field line that surrounds the O point and starts to flow along it.

In Fig. 4 the electron trajectory is traced for a low $\beta$
plasma, with a strong guide field, $B_{y0}=5B_0$. Initially, the
particle flows along the magnetic line, which is mainly directed
along the $y$ direction, and executes a Larmor motion mostly in
the $(x,z)$ plane. The particle then accelerates towards the X
point, crosses magnetic lines in the $(x,z)$ plane, and gains an
out-of-plane velocity which increases its kinetic energy (see Fig.
5). The particle still couples to the magnetic field and executes
a Larmor orbit around the guide field. Meandering orbits are not
present. In contrast to the case with $B_{y0}=0$, the electron
maintains its $y$ velocity even when far from the reconnection
region because now the gyration is in the $(x,z)$ plane around the
$y$-directed guide field. Finally, the electron drifts along a
magnetic field line around the O point maintaining a still
significant $y$ velocity which decreases slowly because of the
interactions of the electron with the non-drifting plasma
background.

The presence of the guide field changes the nature of electron
acceleration: without guide field, the $y$ velocity is lost while
in presence of guide field is retained even far from the
reconnection region \cite{Horiuchi1997}.

\subsection{Electron fluid velocity}

When $B_{y0}=0$, both  kinetic simulations and theoretical results
\cite{Ricci2002b, Kuznetsova2000} show that the electrons are
demagnetized at the electron meandering distance [see Eqs.
(\ref{medis})] and have an inflow and outflow velocity given by
Eqs. (\ref{vel}). The scaling laws of the dimensions of the
reconnection region and of the inflow and outflow velocity, based
on the electron pressure as a break-up mechanism as derived in
Ref. \cite{Kuznetsova2000}, have been verified up to the physical
mass ratio \cite{Ricci2002b}. In the presence of a guide field,
new components of the $\mathbf{E} \times \mathbf{B}$ field arise,
and the electron in-plane motion has been described in Refs.
\cite{Ricci2003, Yin2003}.

Here, we focus on the electron out-of-plane velocity. In Fig. 6,
the velocity along the axis $z=0$ is depicted with $B_{y0}=0$ for
three different mass ratios, $m_i/m_e=25, 180, 1836$. The maximum
out-of-plane velocity increases with the mass ratio. We note that
the out-of-plane velocity is sizeable only in the reconnection
region.

The out-of-plane velocity can be estimated for $B_{y0}=0$. The
electron lifetime in the reconnection region, $\tau$, from Eqs.
(\ref{medis}-\ref{vel}) is approximately

\begin{equation}
\tau \approx d_{xe} / v_{xe} + d_{ze} / v_{ze} \propto \left(
\frac{m_e^2 T_e^2}{e^4 E_y^4} \right)^{1/4} \label{tauu}
\end{equation}
As the magnetic field is negligible in the reconnection region,
the electrons are freely accelerated and the out-of-plane velocity
can be estimated as
\begin{equation}
v_y \approx \frac{e}{m_e} E_y \tau
\end{equation}
and, using Eq. (\ref{tauu}), it follows that
\begin{equation}
v_y \propto v_{th,e}
\label{sl}
\end{equation}

Since the temperature of the electrons is the same for all mass
ratios, it follows that the electron out-of-plane velocity scales
with $1/\sqrt{m_e}$. The results presented in Fig. 6 fit well this
scaling law, as is shown in Table I.

Figures 7 and 8 consider the effect of the guide field on the
out-of-plane electron velocity. As the guide field allows the
particles to flow more easily in the out-of-plane direction, the
peak velocity increases remarkably when the guide field becomes
stronger, Fig. 7. Moreover, the presence of the guide field
changes the general pattern of the out-of-plane velocity, as shown
in Fig. 8. When $B_{y0}=0$, the out-of-plane velocity is sizeable
only near the reconnection region, where the electrons are
accelerated by the electric field. The out-of-plane velocity is
lost when the electrons become again magnetized and are diverted
by the $B_z$ field. In presence of the guide field, the electrons
maintain their $y$ velocity when they leave the reconnection
region and orbit around the O point. Note that this conclusion is
further supported by the analysis of particle orbits performed in
the subsection above (III.b).

As a final remark, we note that the out-of-plane velocity evolves
during magnetic reconnection, as is shown in Fig. 9. For all the
guide fields considered, the electron velocity increases while
reconnection proceeds (the evolution of the reconnected flux in
these simulations is presented in Fig. 1). After reconnection
saturates, the reconnection electric field vanishes and the
electrons are no longer accelerated.  The out-of-plane velocity in
the former reconnection region decreases abruptly.

\subsection{Electron temperature}

The evolution of the electron temperatures in the reconnection
region, $T_{xe}$, $T_{ye}$, and $T_{ze}$, are plotted in Fig. 10
for the three different guide field strengths. We note that
$T_{xe}$, $T_{ye}$, and $T_{ze}$ are defined as the second moment
of the distribution functions of the $x$, $y$, and $z$ velocity
\cite{Cercignani1988}.

In the zero guide field case, the $T_{xe}$ and $T_{ye}$ evolution
is similar (see Fig. 10a). $T_{xe}$ increases because the positive
and negative outflow velocity causes an increases spread in the
$x$ velocity. The heating in the $y$ direction is due to the
electric field which, besides accelerating the electrons, spreads
out their velocity, reflecting the variation in electron residence
time in the reconnection region, which depends on their in-plane
inflow and outflow velocity, and thus are accelerated by different
amounts. After reconnection saturates, the heating process stops
and electrons tend to thermalize, causing an increase in $T_{ze}$.
We note that the total energy of the system is conserved during
the simulation within an error of the order of 4\%
\cite{Ricci2003}.

When the guide field is introduced, both $T_{xe}$ and $T_{ze}$
remain almost constant at the initial level during the
reconnection process, while $T_{ye}$ increases remarkably. The
guide field introduces an higher electron mobility in the $y$
direction. Thus, the electron can be accelerated by the electric
field more than in the $B_{y0}=0$ case along the $y$ direction,
and the $y$ velocities spread out more, while $T_{ye}$ increases.

The anisotropy in the electron temperature contributes to the
break-up of the frozen-in condition. In fact, in the presence of a
guide field, the difference between the diagonal terms of the
electron pressure tensor contributes to the off-diagonal terms,
which are responsible for the break-up of the frozen-in condition
\cite{Ricci2003, Hesse2002}, as it is \cite{Hesse2002}
\begin{equation}
P_{xye}=-\frac{P_{zze}}{\omega_{ce}} \frac{\partial v_{ye}}
{\partial z} + \frac{B_x}{B_y} (P_{yye}-P_{zze})
\end{equation}
\begin{equation}
P_{yze}=-\frac{P_{xxe}}{\omega_{ce}} \frac{\partial v_{ye}}
{\partial x} + \frac{B_z}{B_y} (P_{yye}-P_{xxe})
\end{equation}

Since $P_{yye}-P_{zze} = n_e(T_{ye}-T_{ze})$ and $P_{yye} -P_{zze}
= n_e (T_{ye}-T_{ze})$, the importance of the anisotropy in the
electron temperature is evident.

\section{CONCLUSIONS}

In the present paper, the electron dynamics during magnetic
reconnection has been studied by showing and discussing results of
kinetic simulations of Harris current sheets. Simulations with
different plasma $\beta$ and different mass ratio have been
considered.

By varying the guide field, we have been able to model
reconnection in systems such as the magnetotail ($B_{y0}=0$), the
magnetopause ($B_{y0}=B_0$), laboratory and astrophysical plasmas
($B_{y0}=5 B_0$).

By studying the typical electron trajectories, we have shown that,
when the plasma $\beta$ is decreased, the electrons mainly perform
Larmor motion around the guide field even in the reconnection
region, and that meandering orbits disappear. In all the cases,
electrons are accelerated by the reconnection electric field along
the $y$ direction and their velocity increases with the guide
field. Moreover, the out-of-plane velocity increases during
reconnection. In high $\beta$ plasmas, the out-of-plane velocity
is sizable only in the electron reconnection region. With a guide
field, the out-of-plane velocity is globally relevant. The mass
ratio has a strong influence on the out-of-plane velocity and the
scaling law of interest is derived. The study of the electron
temperature in the reconnection region has shown a strong heating
anisotropy in presence of a guide field, which contributes to the
break-up of the electron frozen-in condition.

In closing, we note that we plan to develop the present work in
two directions. First, we plan to introduce the relativistic
equations of motion in CELESTE3D, in order to represent better the
electron physics when relativistic effects become important.
Second, an experimental setup has been built at the Los Alamos
National Laboratory to study reconnection experimentally in
plasmas with different $\beta$ \cite{Furno2003}. We plan to
compare our simulation results with the experiments.

\section*{ACKNOWLEDGMENTS}

This research is supported by the LDRD program at the Los Alamos
National Laboratory, by the United States Department of Energy,
under Contract No. W-7405-ENG-36 and by NASA, under the "Sun Earth
Connection Theory Program". The supercomputer used in this
investigation was provided by funding from the JPL Institutional
Computing and Information Services and the NASA Offices of Space
Science and Earth Science.

\newpage

\begin{itemize}

\item

Fig. 1 (from Ref. \cite{Ricci2003}): Reconnected flux (normalized
to $B_0c/\omega_{pi}$), for $m_i/m_e=25$ (a), $m_i/m_e=180$ (b),
and $m_i/m_e=1836$ (c), and $B_{y0}=0$ (solid line), $B_{y0}=B_0$
(dashed line), $B_{y0}=5B_0$ (dotted line).

\item

Fig. 2: Electron trajectory in the $(x,z)$ plane, for $B_{y0}=0$
and $m_i/m_e=25$. The position of the particle at different times
is marked by circles, the starting position by a plus sign. The
position of the X point is denoted by the x-mark. Note the
periodic boundary conditions in the $x$ direction.

\item

Fig. 3: Velocities $v_x$ (a), $v_y$ (b), $v_z$ (c) and kinetic
energy (d), as a function of time, for the electron whose
trajectory is represented in Fig. 2.

\item

Fig. 4: Electron trajectory in the $(x,z)$ plane, for $B_{y0}=5
B_0$ and $m_i/m_e=25$. The position of the particle at different
times is marked by circles, the starting position by a plus sign.
The position of the X point is denoted by the x-mark. Note the
periodic boundary conditions in the $x$ direction.

\item

Fig. 5: Velocities $v_x$ (a), $v_y$ (b), $v_z$ (c) and kinetic
energy (d), as a function of time, for the electron whose
trajectory is represented in Fig. 4.

\item

Fig. 6: Electron out-of-plane velocity at $z=0$, when $\Delta \Psi
\approx 1$, for the simulations with $B_{y0}=0$ and $m_i/m_e=25$
(dashed line), $m_i/m_e=180$ (dotted line), and $m_i/m_e=1836$
(solid line).

\item

Fig. 7: Electron out-of-plane velocity at $z=0$, when $\Delta \Psi
\approx 1$, for the simulations with $B_{y0}=B_0$ (a) and
$B_{y0}=5B_0$ (b), $m_i/m_e=25$ (dashed line), $m_i/m_e=180$
(dotted line), and $m_i/m_e=1836$ (solid line).

\item

Fig. 8: Electron out-of-plane velocity when $\Delta \Psi \approx
1$, for the simulations with $B_{y0}=B_0$ (a) and $B_{y0}=5B_0$
(b), for the simulations with $m_i/m_e=180$.

\item

Fig. 9: Evolution of the average out-of-plane electron velocity,
$v_y$ in the reconnection region, for the simulation with
$m_i/m_e=25$, and guide field $B_{y0}=0$ (a), $B_{y0}=B_0$ (b),
and $B_{y0}=5B_0$ (c).

\item

Fig. 10: Evolution of the electron thermal velocity $v_{thx,e}$
(solid), $v_{thy,e}$ (dashed), and $v_{thz,e}$ (dotted), in the
reconnection region for the simulation with $m_i/m_e=25$ and
$B_{y0}=0$ (a) and $B_{y0}=5B_0$ (b).

\end{itemize}

\bigskip

\textbf{Table I.}Comparison between the simulation results and the
scaling law in Eq. (\ref{sl})

\begin{tabular}
[c]{lcc} \hline\hline
Ratio & Simulation result & Scaling law\\
\hline
$\frac{v_y(m_i/m_e=180)}{v_y(m_i/m_e=25)}$ & 2.9 & 2.7\\
$\frac{v_y(m_i/m_e=1836}{v_y(m_i/m_e=25)}$ & 7.0 & 8.6\\
\hline\hline
\end{tabular}

\begin{figure}[p]
\centering
\includegraphics[]{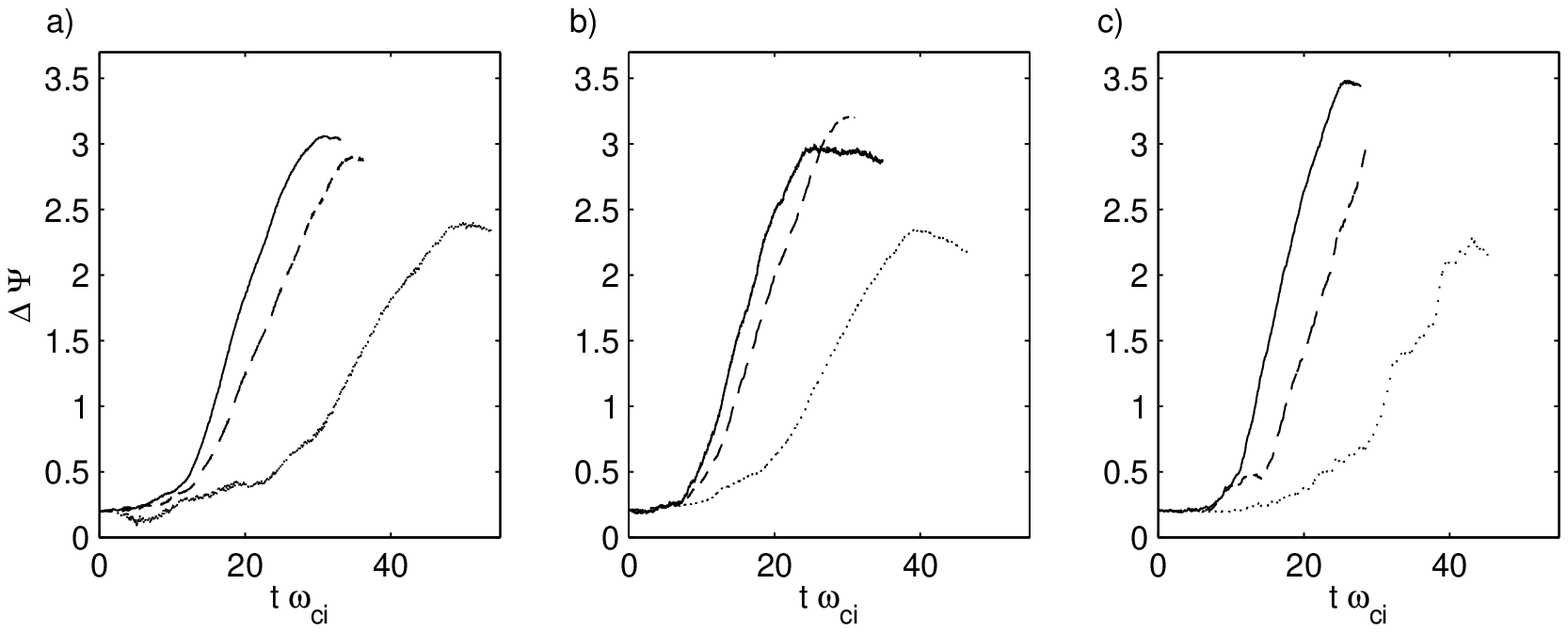}
\caption{}
\end{figure}

\begin{figure}[p]
\centering
\includegraphics[]{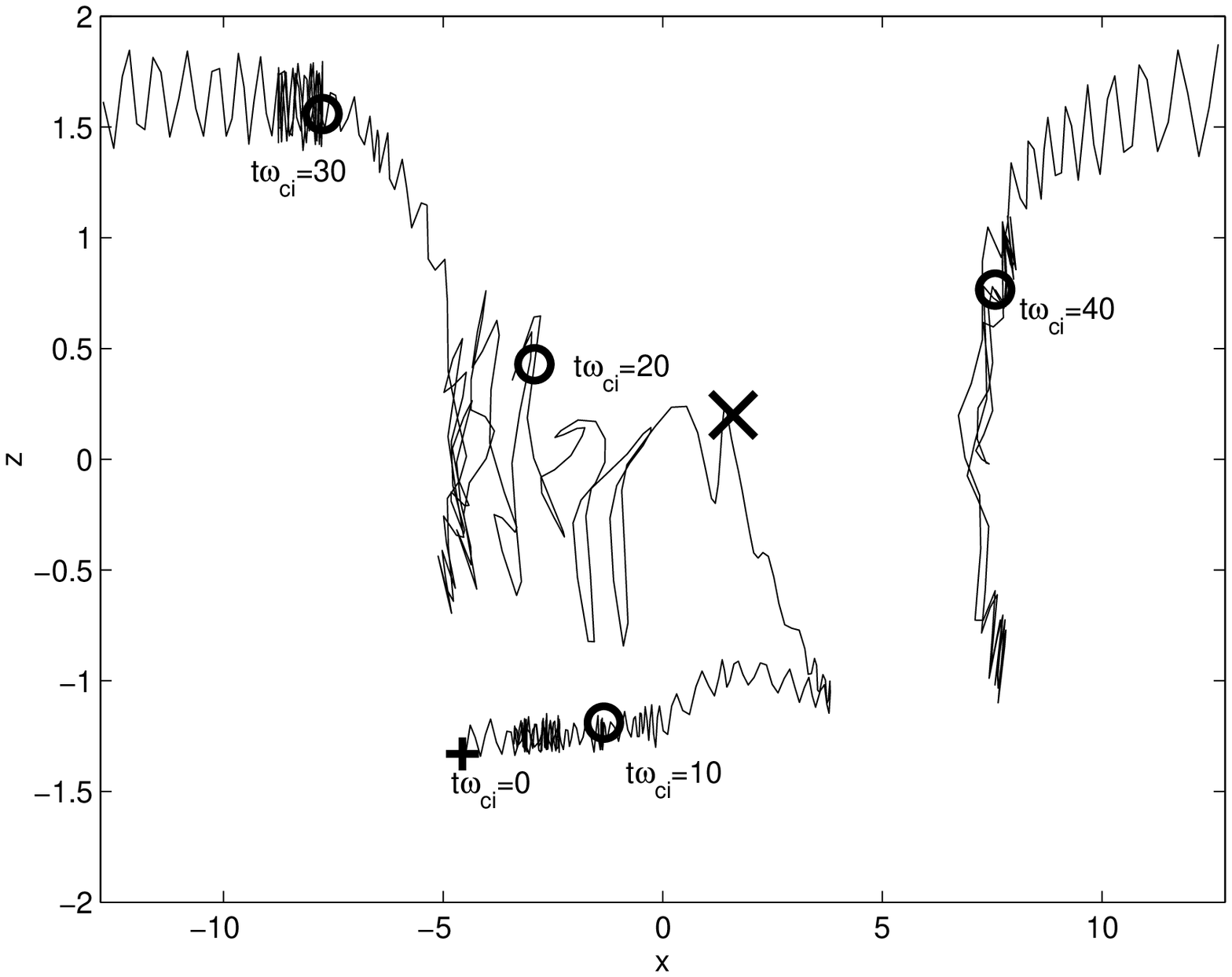}
\caption{}
\end{figure}

\begin{figure}[p]
\centering
\includegraphics[]{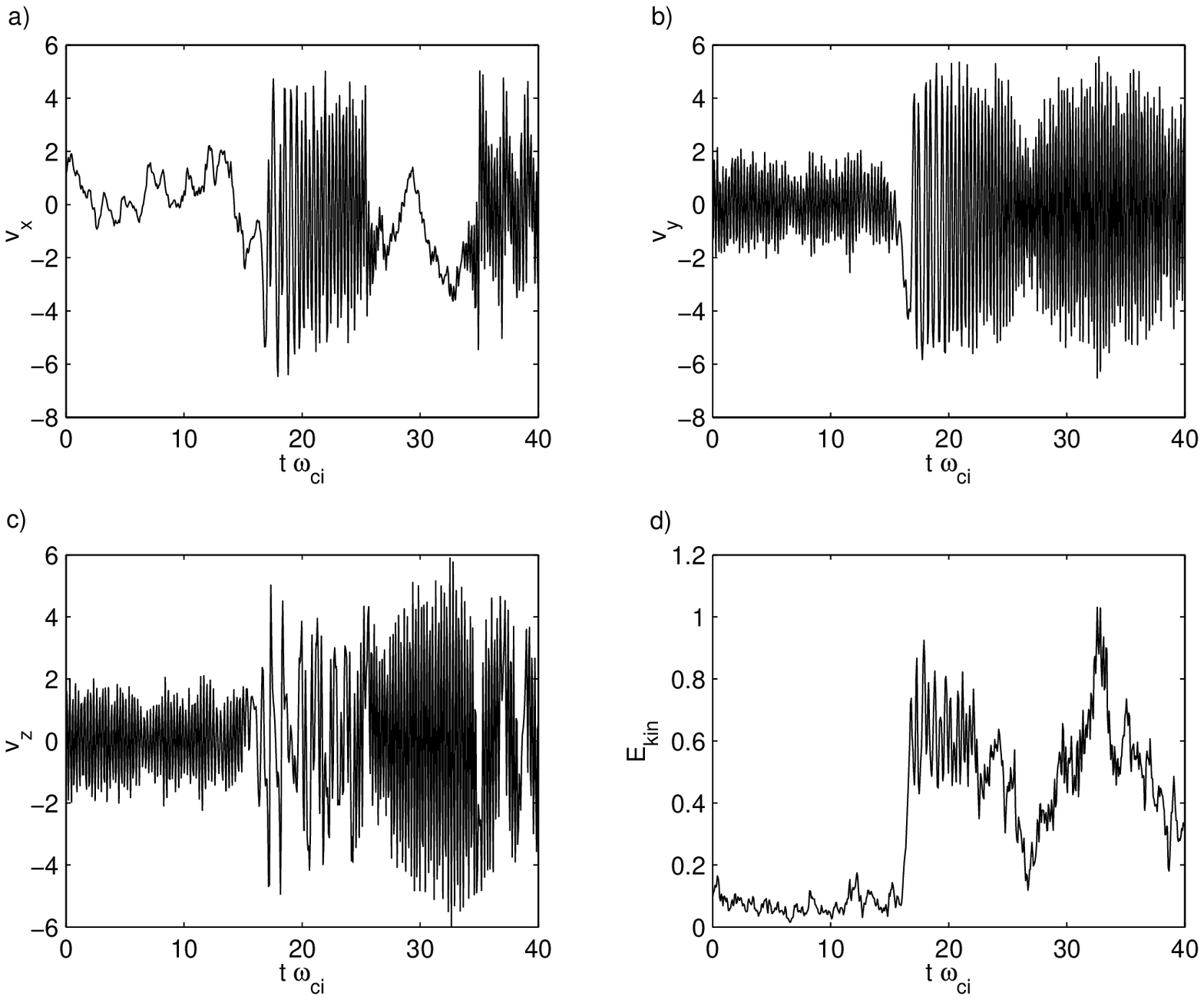}
\caption{}
\end{figure}

\begin{figure}[p]
\centering
\includegraphics[]{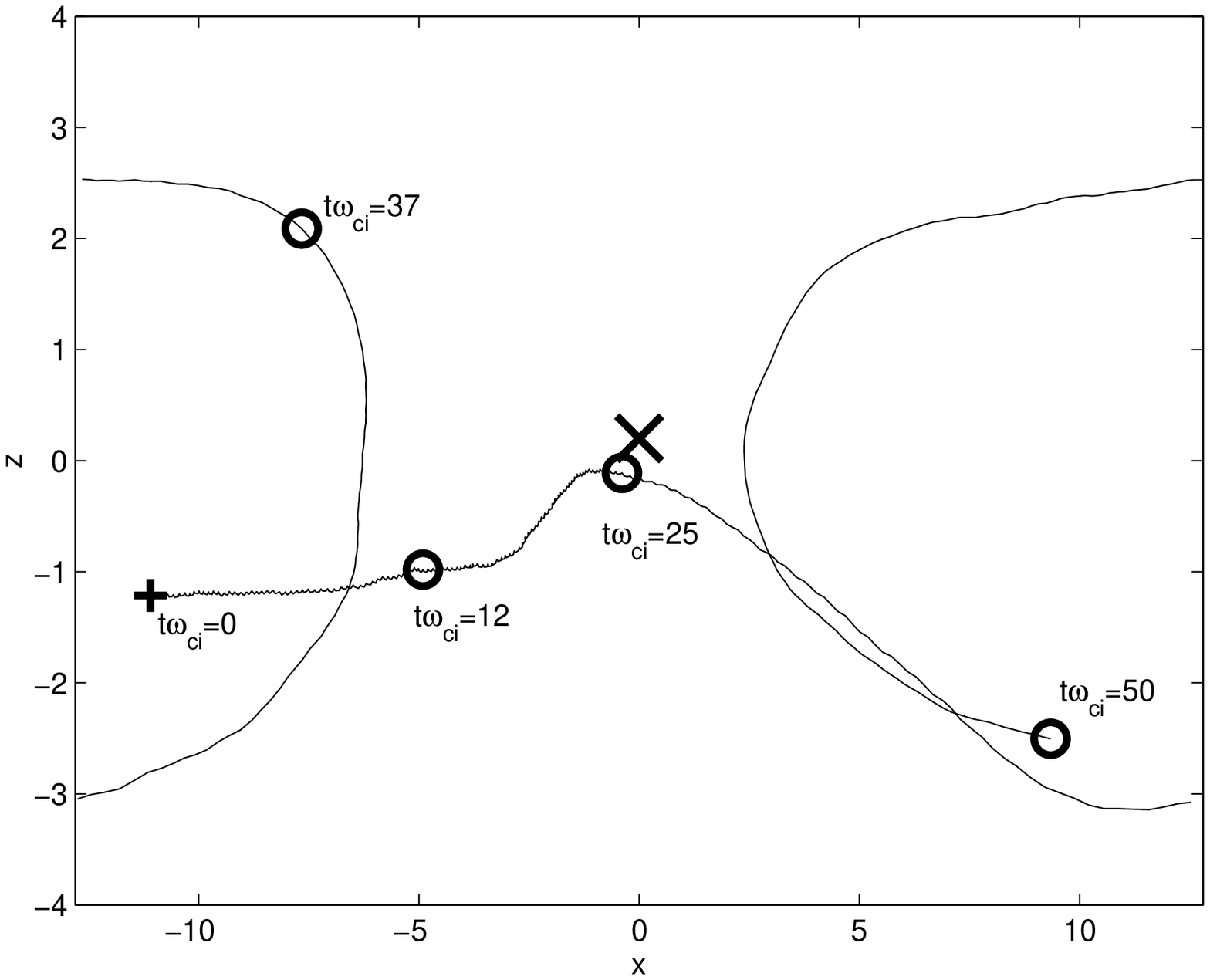}
\caption{}
\end{figure}

\begin{figure}[p]
\centering
\includegraphics[]{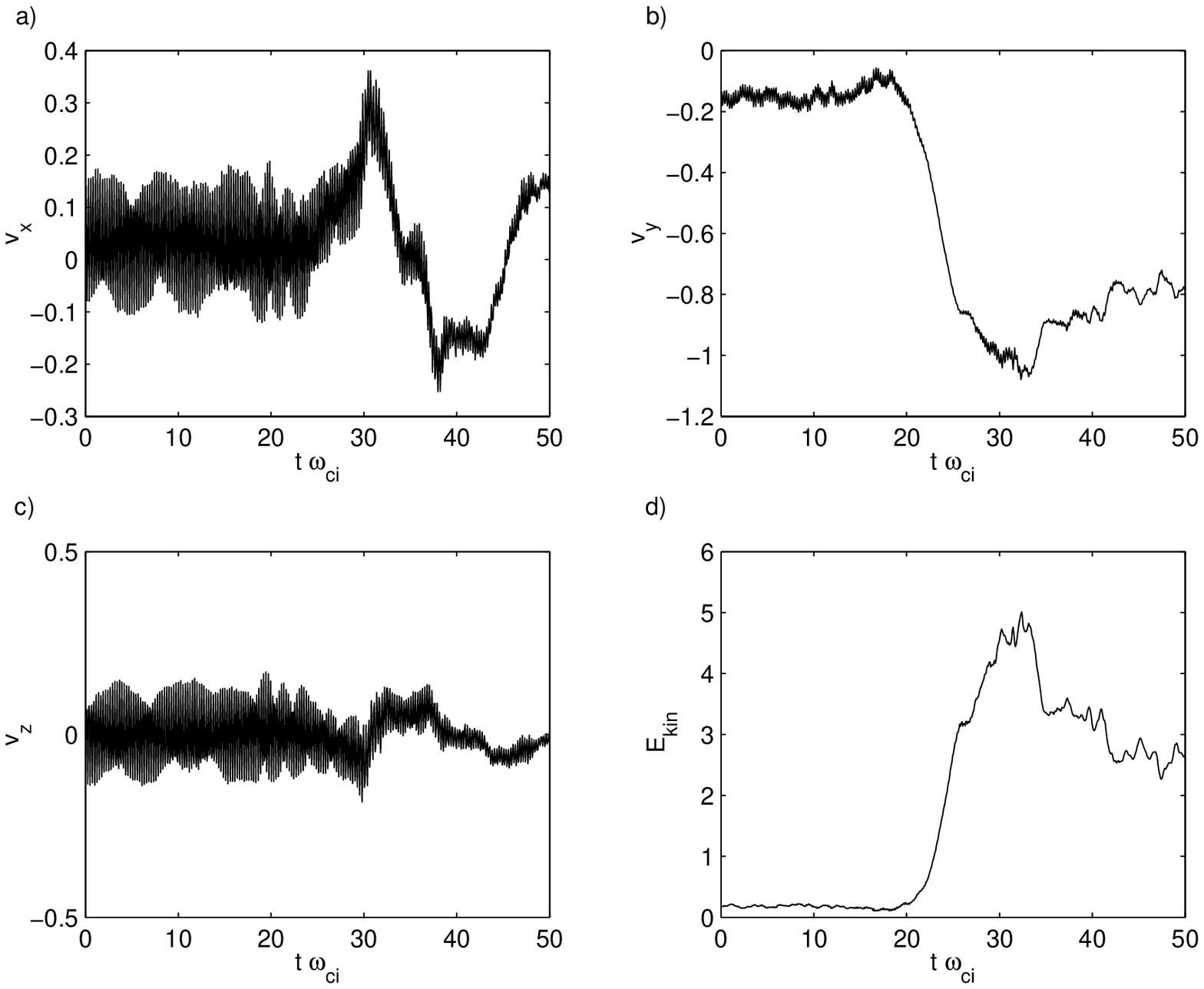}
\caption{}
\end{figure}

\begin{figure}[p]
\centering
\includegraphics[]{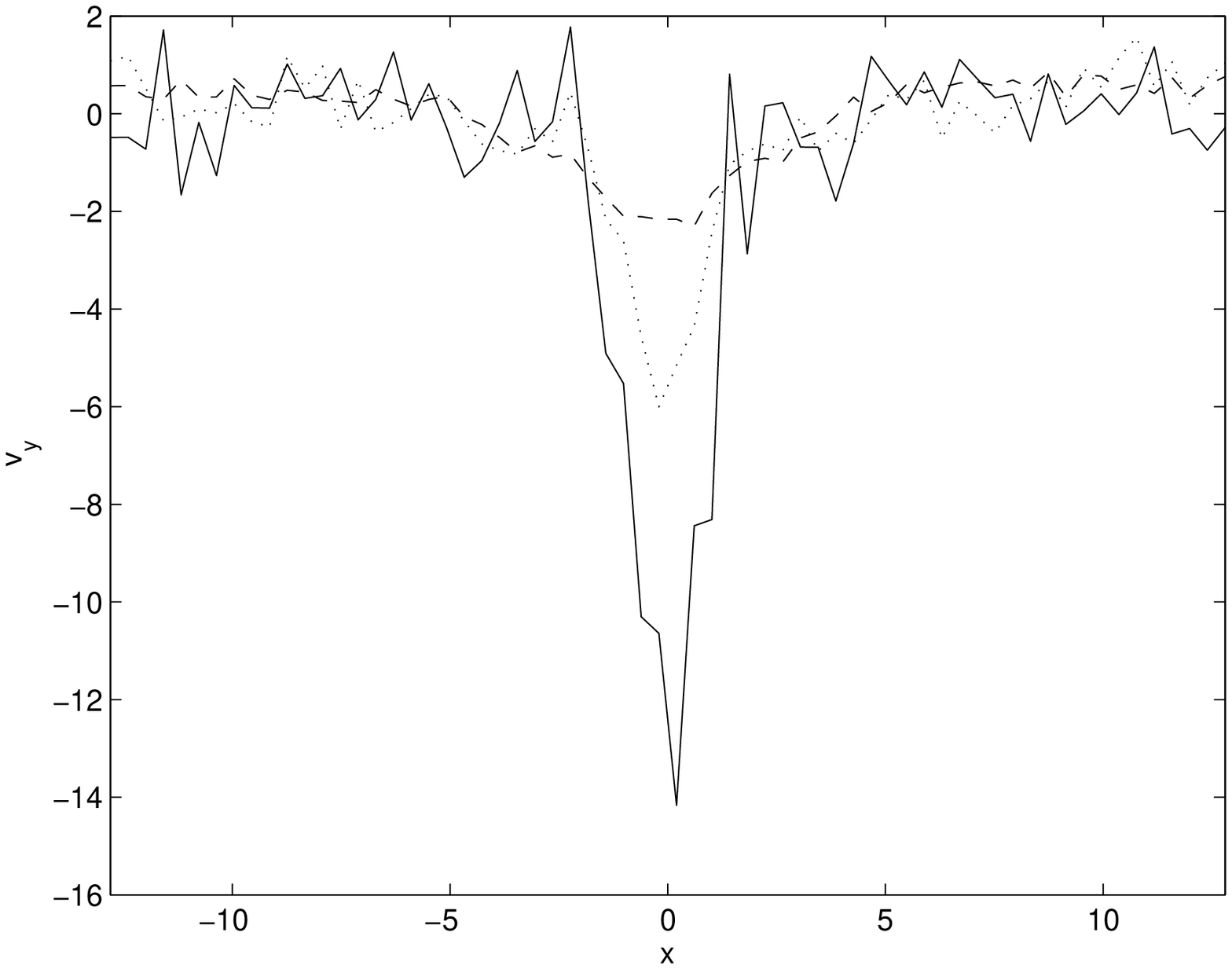}
\caption{}
\end{figure}

\begin{figure}[p]
\centering
\includegraphics[]{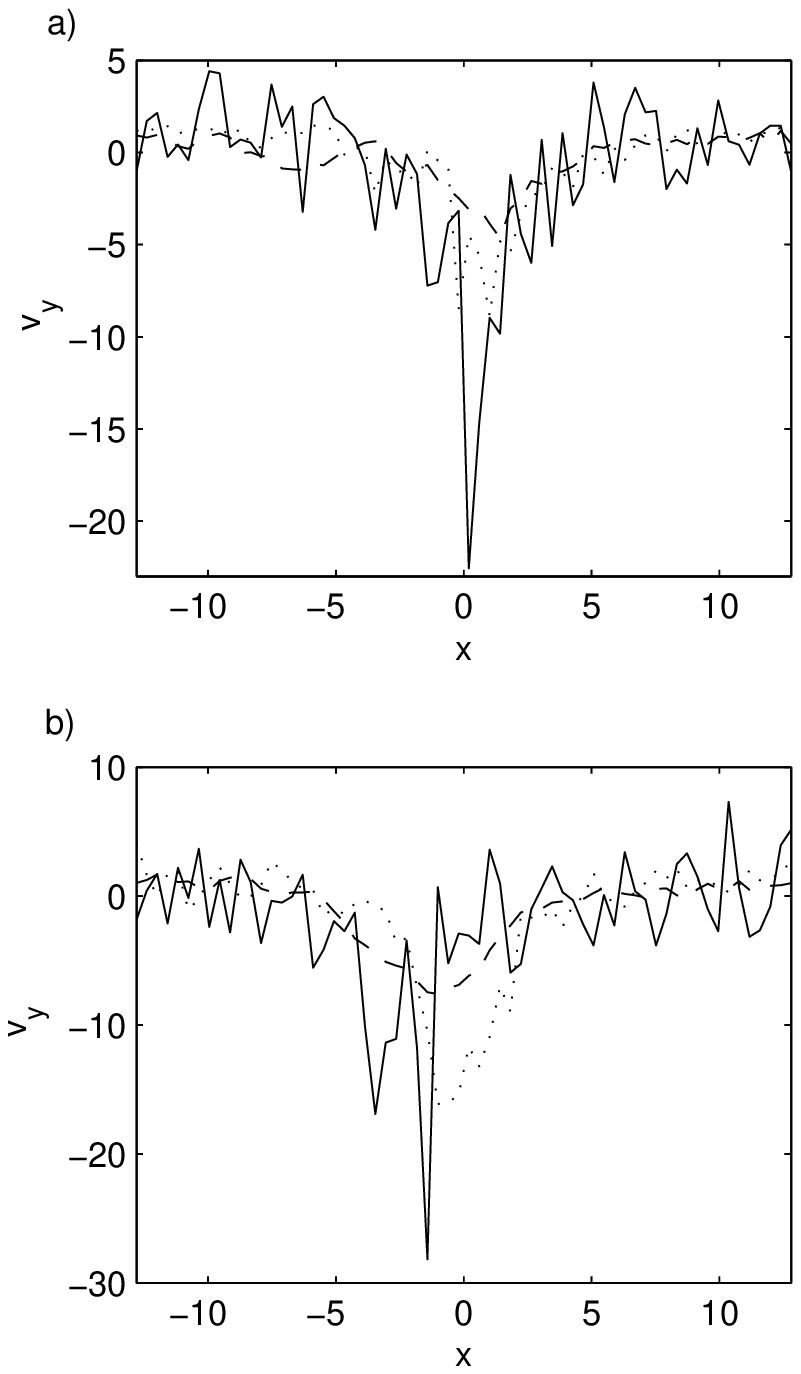}
\caption{}
\end{figure}

\begin{figure}[p]
\centering
\includegraphics[]{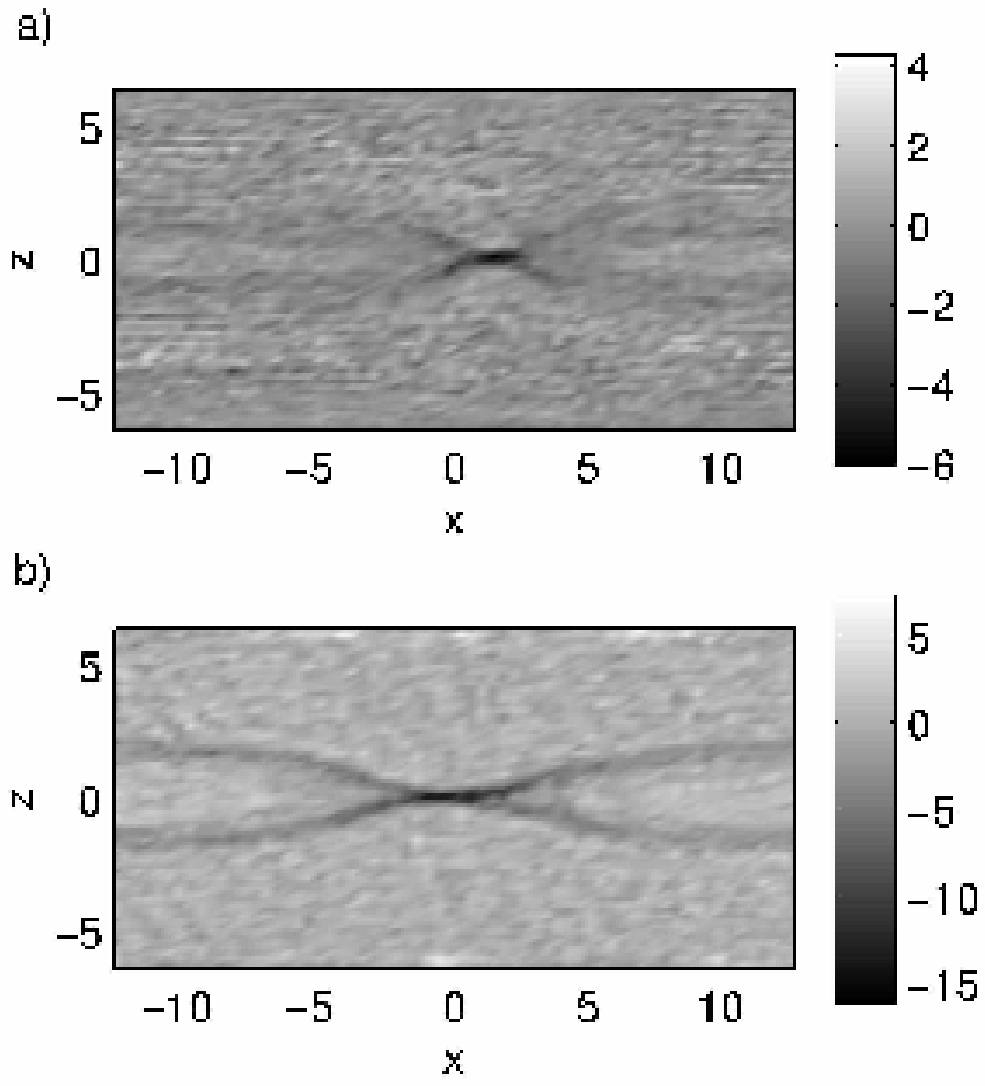}
\caption{}
\end{figure}

\begin{figure}[p]
\centering
\includegraphics[]{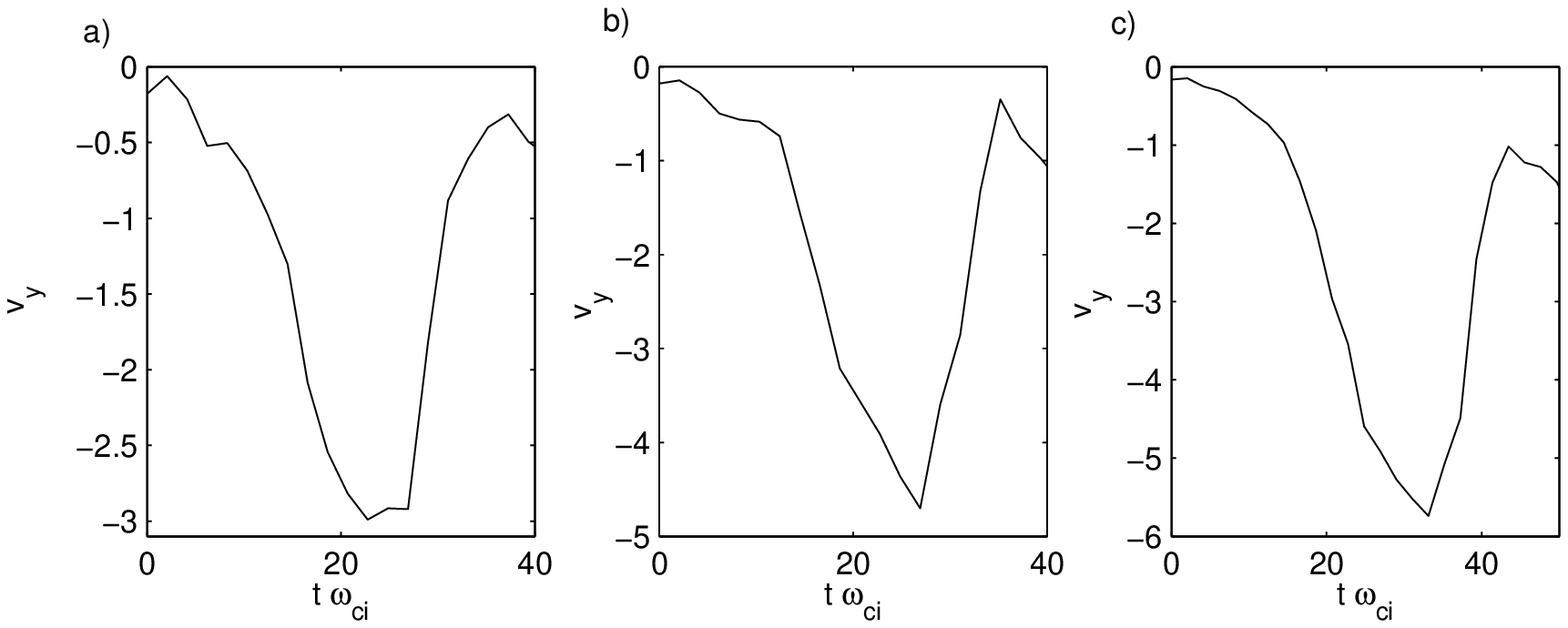}
\caption{}
\end{figure}

\begin{figure}[p]
\centering
\includegraphics[]{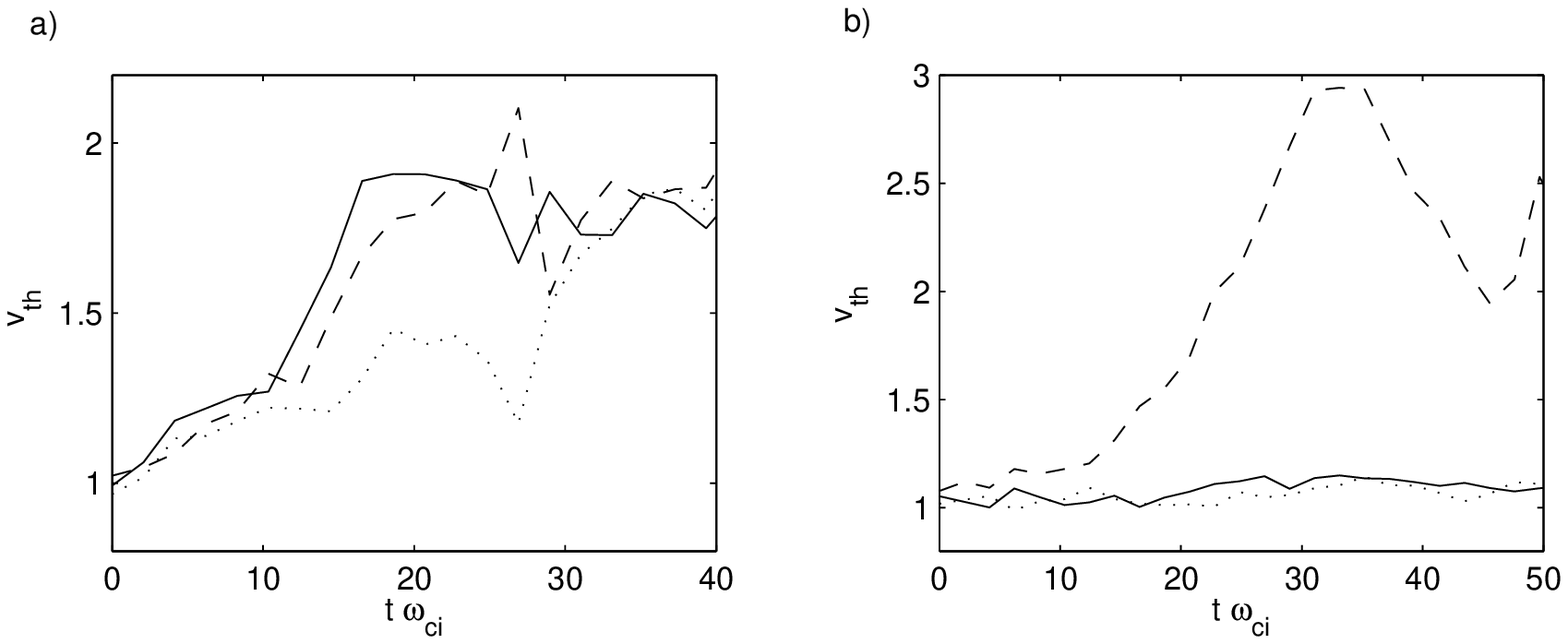}
\caption{}
\end{figure}

\end{document}